\documentclass[doublecol]{epl2}
\usepackage{epsf}
\usepackage{amsfonts}		
\usepackage{amsmath}
\usepackage{color}
\usepackage{multirow}
\definecolor{c1}{rgb}{1, 0, 0}
\definecolor{c2}{rgb}{0, 1, 0}
\definecolor{c3}{rgb}{0, 0, 1}
\definecolor{c4}{rgb}{1, 0, 1}
\definecolor{c5}{rgb}{0, 1, 1}

\def\ie{\hbox{\it i.e.}}
\def\etal{\hbox{\it et al.}}
\def\nn{\nonumber}
\def\beq{\begin{equation}}
\def\eeq{\end{equation}}
\def\bea{\begin{eqnarray}}
\def\eea{\end{eqnarray}}

\title{Influence of long-range interactions on the critical behavior of the Ising model}

\author{T. Blanchard\inst{1}, M. Picco\inst{1} and M. A. Rajapbour\inst{2,3}}
\shortauthor{T. Blanchard, M. Picco and M. A. Rajapbour}

\institute{
	\inst{1} CNRS, LPTHE, Universit\'e Pierre et Marie Curie, UMR 7589 - 4 place Jussieu, 75252 Paris cedex 05, France, EU\\
	\inst{2} SISSA and INFN, Sezione di Trieste - via Bonomea 265, 34136 Trieste, Italy, EU\\
	\inst{3} Instituto de F\'{\i}sica de S\~{a}o Carlos, Universidade de S\~{a}o Paulo, Caixa Postal 369, 13560-590, S\~{a}o Carlos, SP, Brazil
}


\date{\today}

\abstract{
We study the ferromagnetic Ising model with long-range interactions in two dimensions. We first present results of a
Monte Carlo study which shows that the long-range interactions dominate over the short-range ones in the intermediate
regime of interaction range. Based on a renormalization group analysis, we propose a way of computing the influence of
the long-range interactions as a dimensional change.
}

\begin{document}
\maketitle

\section{Introduction}  
In recent years there has been a lot of interest in the statistical physics of classical and quantum systems with
long-range interactions, for a review see~\cite{Campa}.  The role of quasi-stationary states and ergodicity breaking in
long-range interacting systems was investigated in~\cite{Gabrielli} and~\cite{Benetti}. In~\cite{Kastner} the
approaching to equilibrium for long-range quantum systems was examined and there has been a lot of
enthusiasm in investigating the entanglement entropy in long-range spin chains~\cite{Barthel,Koffel,Cadarso}. Very
recently an experiment was conducted on a quantum system with tunable long-range interactions~\cite{Britton}.

In the present study we focus on the Ising model which is probably the most studied model in statistical mechanics,
especially in the context of critical phenomena. Most of the studies about the  Ising model are concentrated around the
short-range case which is exactly solvable in one  and two dimensions~\cite{Baxter}. In three dimensions the problem was
perturbatively studied using the $\epsilon$-expansion technique~\cite{wilsonFisher} of the renormalization group (RG)
combined with the Borel resummation of the perturbation series, see~\cite{RGuida}  and references therein. Most recently
the problem was revisited by using conformal bootstrap technique~\cite{Rychkov}. Although now there are little unknown
facts around short-range Ising model the  long-range Ising model  is still the subject of many contradicting theoretical
and numerical studies. We define the long-range Ising model as 
\beq
{\cal H} = -  \sum_{\langle ij\rangle} \frac{J}{r_{ij}^{d+\sigma}} S_i S_j   \; ,
\eeq
where the sum is over all pairs of spins of a $d$ dimensional system and $J>0$ . In~\cite{fisher}
the $\epsilon$-expansion technique was applied  to the above problem shortly after the introduction of the method. Three
regimes were discovered: (a) the classical regime $0<\sigma<d/2$ with mean-field critical exponents; (b) the intermediate
regime $d/2<\sigma<2$, where the exponents are functions of $\sigma$ and (c) the short-range regime $\sigma>2$ where the
exponents are the same as in the short-range Ising model. The conjectures around the first regime are already proved
in~\cite{Aizenman} and  the results of the third regime are widely accepted. The intermediate regime has been the subject
of many controversies in the last forty years. 
In~\cite{fisher} Fisher \etal obtained the expression for $\eta$ and $\gamma$ the susceptibility exponent up
to $\epsilon'^2$ for $\sigma<2$ with $\epsilon'=2\sigma-d$. They observed a discrepancy for both exponents at this order
at $\sigma=2$ between their expression for $\sigma<2$ and their short-range value for $\sigma>2$. The case of the
exponent $\eta$ is special because it gets no corrections at this order so that it sticks to its classical regime's
value; \ie $\,\,\eta=2-\sigma$.  Shortly after, Sak~\cite{sak} argued that there is no such discontinuity for
$\eta$, $\gamma$ and $\varphi$ the crossover exponent because if one looks at the long-range interaction as a perturbation
of the short-range Ising, $\sigma=2$, one can discover that the short-range Ising exponents should be extended
to $\sigma=2-\eta_{\,\mathrm{\textsc{sr}}}$ and so there is no discontinuity in the critical exponents. 

Although  some other forms of RG also
appeared~\cite{Yamazaki} in the last forty years, Sak's argument has been widely accepted~\cite{Cardy}.
An especially interesting numerical simulation done by Bl\"ote and Luijten~\cite{BL} completely ruled out any jump in the value of $\eta$. 
The motivation of our work comes from a recent numerical work done by one of us~\cite{mp} in which we improved the algorithm 
of Bl\"ote and Luijten. This algorithm uses the fact that for a ferromagnetic model, one can use clusters of spins 
to  improve the speed of the simulations as is done in the Wolff cluster algorithm for short range ferromagnetic models~\cite{W}. 
The improvement in~\cite{mp} concerns the construction of the clusters by optimizing the search of connected spins over large regions. 
With this new algorithm, we can simulate systems up to size $5120 \times 5120$ with a typical update time of order one second 
on an ordinary workstation. By analyzing much bigger sizes than in previous studies, we concluded
that neither Fisher \etal's procedure nor Sak's machinery fit with the numerics, in particular in the 
intermediate regime and close to the boundary with the short-range regime. 

In the present study, we will concentrate on two aspects of this problem. First, we will compare the long-range behavior
with the short-range one in two dimensions. We provide numerical evidences that the long-range interactions dominate for $\sigma \leq 2$.
Next, we will propose another way to compute the $\eta$ exponent. The main idea is to make a correspondence between
\begin{equation}
\label{t1}
A_1 = \int\!\! \mathrm{d}^dx \left( \frac{1}{2}  |\nabla^{\sigma/2} s_b(x)|^2 +\frac{\lambda_0}{4!} |s_b(x)|^4  \right),
\end{equation}
with $\epsilon'=2\sigma-d$  and
\begin{equation}
\label{t2}
A_2 = \int\!\! \mathrm{d}^Dx \left( \frac{1}{2}  |\nabla s_b(x)|^2 +\frac{\lambda_0}{4!} |s_b(x)|^4  \right),
\end{equation}
with $\epsilon=4-D$.   
The first expression $A_1$ is a formal way of writing in real space a model with long-range interactions. 
The second expression $A_2$ is the usual way of expressing a short-range $\phi^4$ model for a $D$-dimensional theory. 
For $\sigma \simeq 2$ and with the condition $\epsilon = \epsilon'$,
the $\epsilon$-expansion of both models, $A_1$ and $A_2$ will be the same apart from a term proportional
to $\delta \sigma = 2-\sigma$. Thus the computation will be done
from the model $A_2$ with $D=4+d-2\sigma $. The deviation of $\sigma$ from $2$ is
replaced by a deformation of the dimension from $d$ to $D$.  

\section{Long Range versus Short Range}
In \cite{mp}, it was observed that the behavior of the model with long-range interactions and for $\sigma < 2$ is different
from what is expected for the short-range model. Then we must worry about the relevance of the long-range
interactions compared to the short-range ones.
If we start from a short-range model and consider the addition of long-range term 
$g \sum_{ij} S_i S_j / r_{ij}^{d+\sigma}$ 
as a small perturbation, then a simple dimensional argument predicts  
the relevance of the perturbation as a function of the dimension of $g$~\cite{Cardy}.
Since for large distances we have $\left< S_i S_j \right> \simeq r_{ij}^{2-d - \eta_{\,\mathrm{\textsc{sr}}}}$, with $\eta_{\,\mathrm{\textsc{sr}}}=\frac{1}{4}$ in two dimensions,
we  expect that the dimension of $g$ is $2 - \eta_{\,\mathrm{\textsc{sr}}} - \sigma$.  Then the result is that the perturbation is relevant
for $\sigma < 2 - \eta_{\,\mathrm{\textsc{sr}}}$ and irrelevant otherwise.
\begin{figure}
\epsfxsize=240pt{\epsffile{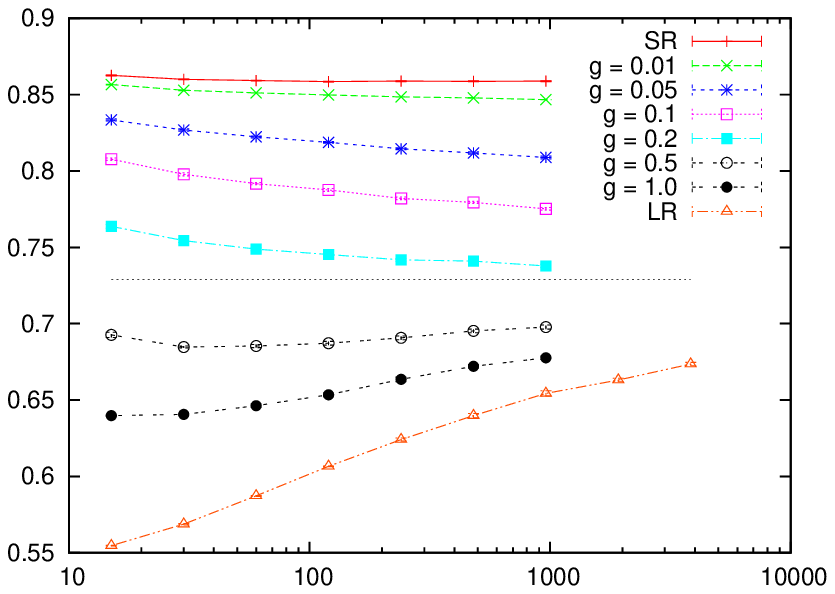}}
\epsfxsize=240pt{\epsffile{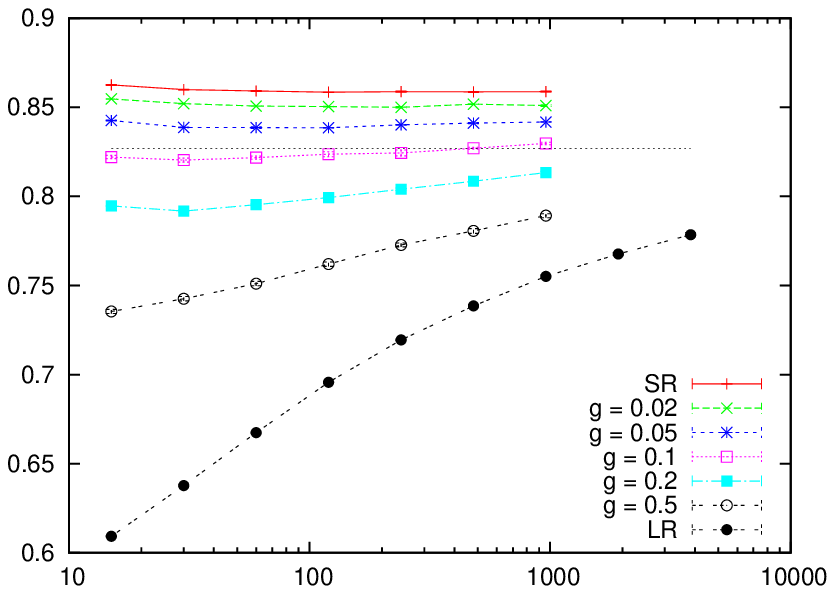}}
\caption{(Color online)  $B(g,L)$ vs. $L$ for $\sigma=1.6$ (top panel) and $\sigma=1.8$ (bottom panel). 
The short-range (SR) results are depicted in red and have the largest $B(g,L)$. The dotted line corresponds to the 
extrapolated value for the long-range (LR) model (see~\cite{mp}).
}
\label{PB}
\end{figure}
We will now test this argument. We  consider the case of the perturbation as a  function of $\sigma$ and $g$.
We will use the magnetic cumulant defined as: 
\begin{equation}
B(g,L,K)=\frac{\langle m^2\rangle^2}{\langle m^4\rangle},
\label{magc}
\end{equation}
where $K=\beta J$. For each value of $g$, we consider the quantity $B_c(g,L,L')$ which corresponds to the crossing of
$B(g,L,K)$ and $B(g,L',K)$ as a function of $K$.  By choosing a set of increasing values $L$ and $L'$ not too far apart, we can determine
for each pair the value of $K$ which corresponds to the crossing and $B_c(g,L,L')$ is expected to converge towards
a finite limit for $L \rightarrow +\infty$.
In the following, we will always consider $L'=2 L$ and then we will just
denote the crossing value by $B(g,L)$.  In \cite{mp}, it was determined that the corresponding quantity for the
long-range interaction model, which can be considered as the limit $g\rightarrow \infty$, converges to a value smaller
than the one of the short-range model for $\sigma \leq 2$. In Fig.~\ref{PB}, we present the measured values of $B(g,L)$
for $\sigma=1.6$ and $\sigma=1.8$.  For the first case, we observe a clear tendency for $B(g,L)$ to converge towards
the same limit as the LR model (which is shown as a dotted line) and this for all the values of $g$ in the range $g=0.01$ up to $g=1$. For the second
case, the situation is less clear.  For the small values of $g \leq 0.1$, it seems first that $B(g,L)$ converges
towards the model with short-range interactions. For larger values of the perturbation, we just  observe that $B(g,L)$
increases with the size. While it can be assumed that this just corresponds to the flow towards the value for the SR
model, one can also invoke the effect of strong finite size corrections. 

\begin{figure}
\epsfxsize=240pt{\epsffile{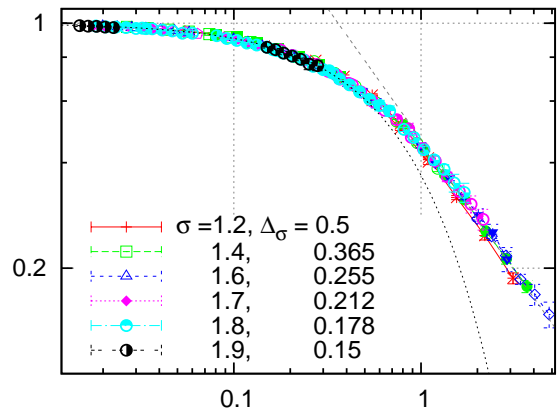}}
\caption{(Color online)  $X(g,L) $ vs. $g L^{\Delta_\sigma}$. The straight dashed line corresponds to $(g
L^{\Delta_\sigma})^{-0.75}$, while the other dashed curve is for $\exp(-gL^{\Delta_\sigma})$.
}
\label{PS}
\end{figure}

In fact, since we are considering a case in which there is
both a flow towards either the LR model or the SR model and  very strong finite size effects, it is difficult to
know which one is the dominant effect. We then adopt another strategy. We will look in the following to the quantity defined as 
\begin{equation}
X(g,L) = \frac{B(g,L) - B_{\,\mathrm{\textsc{lr}}}(L)}{B_{\,\mathrm{\textsc{sr}}}(L)-B_{\,\mathrm{\textsc{lr}}}(L)} \; .
\end{equation}
This quantity is defined such that it takes a value between $0$ and $1$. If $B(g,L)$ flows towards the SR point, then
$X(g,L)$ goes to $1$. On the contrary, if it flows towards the LR point, then it goes to $0$. We then expect the
crossover to be controlled by a crossover parameter $g / |t|^{\phi}$ with $t$ the reduced temperature and $\phi = \nu
\Delta_\sigma$ which defines $\Delta_\sigma$. On a finite lattice of linear size $L$ and at the critical point,
this becomes $g L^{ \Delta_\sigma}$ with the correspondence $t^{-\nu}\! \sim \!\xi\! \sim\! L$.
According to the naive dimensional analysis  made above, $\Delta_\sigma = 2- \sigma -
\eta_{\,\mathrm{\textsc{sr}}}$. In Fig.~\ref{PS}, we show a plot of $X(g,L)$ vs. the crossover parameter for various values of $g$ in
$[0.01,1]$ and
$\sigma$. For each value of $\sigma$, we determined a single parameter $\Delta_\sigma$ which allows to make a scaling
for all the values of $g$ on a single curve. The values that we obtained are reported
in the caption of the figure. It is quite remarkable that the curves for all values of $\sigma$ collapse on a single
curve. We obtain that for $ g L^{ \Delta_\sigma}\!\! \ll 1$, the curves follow an exponential,  \ie $\,\,X(g,L) \simeq
\exp{(-g L^{ \Delta_\sigma})}$.  For $ g L^{ \Delta_\sigma}\!\! \gg 1$, the curves behave as a power law \ie
$\,\,X(g,L) \simeq (g L^{ \Delta_\sigma})^{-\alpha}$ with $\alpha \simeq 0.75$.

A second fact is that the value of $\Delta_\sigma$ does not follow the prediction obtained from the naive dimensional
analysis.  While for small values of $\sigma$, the correspondence between the measured crossover exponent and the
predicted one is acceptable,
this is clearly not the case for larger values of
$\sigma$. And in particular, this exponent does not cancel at $\sigma = 2- \eta_{\,\mathrm{\textsc{sr}}}$.  Note also that the precision on
this exponent is not very good for larger values of $\sigma$ since in that case the denominator of $X(g,L)$ becomes 
small and will cancel in the large size limit for $\sigma=2$.

The conclusion of this analysis is that we observe a clear signal for a crossover between the short-range
interaction  model and the long-range interaction model.  This crossover seems to be present in all the range that we
can consider $1.2 \leq \sigma \leq 1.9$. Of course, such a crossover is expected for small values of $\sigma$, \ie
$\,\,\sigma \leq 2 - \eta_{\,\mathrm{\textsc{sr}}} = 1.75$. We observed that in fact this crossover between the SR model towards the LR
model remains present even for larger values of  $\sigma$, presumably up to $\sigma=2.0$.

\section{Renormalization group approach}
\begin{figure} [htb]
\centering
\epsfxsize=200pt{\epsffile{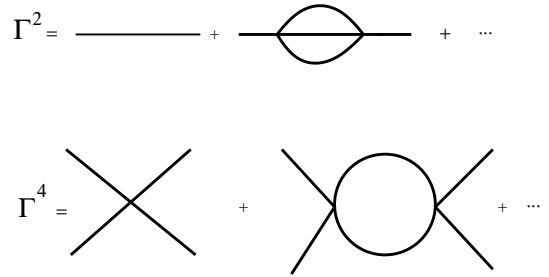}}
\caption{The relevant Feynman diagrams in RG calculation of the beta function and wave function normalization.} \label{Figure:1}
\label{feyn_graphs}
\end{figure}

In this section we propose a new way of doing RG analysis for long-range Ising model. Although our analysis shares some similarities with the work of Yamazaki the final results are more general \cite{Yamazaki}. We implement our RG analysis
around the critical point \cite{Itzykson,Kleinert}, then we can avoid calculating more complicated integrals.  Based on
the arguments of the last section one can write the Lagrangian of the long-range Ising model, forgetting about the
irrelevant  short-range part, with respect to the renormalized coupling and field as
\begin{align}
\label{phi4long-rang Ising}
\mathcal{L}=&\frac{1}{2} |\nabla^{\sigma/2} s(x)|^2+\frac{\lambda}{4!}\mu^{\epsilon'}|s(x)|^4+(Z-1)\frac{1}{2} |\nabla^{\sigma/2} s(x)|^2\nn \\
&+\frac{\lambda}{4!} \mu^{\epsilon'}(Z^2 Z_{\lambda}-1)|s(x)|^4+\dotsb,
\end{align}
where $\epsilon'=2\sigma-d$, $s=Z^{-1/2}s_b$ and $\lambda_{0}=\lambda \mu^{\epsilon'}Z_{\lambda}$ with $s_b$ and $\lambda_0$ as  bare parameters. 
The scale $\mu$ is introduced because we want to do the expansion around the massless theory \cite{Itzykson}.
The third and fourth terms are designed to remove divergent contributions in vertex functions. The renormalization conditions are
\begin{align}
\label{RG conditions}
\Gamma^{(2)}(0)&=0;\\
\frac{\partial}{\partial |p|^\sigma}\Gamma^{(2)}(p,-p)|_{_{|p|^\sigma=\mu^\sigma}}&=1;\\
\Gamma^{(4)}(p_1,p_2,p_3,p_4)|_{_{S_{\mu}}}&=\lambda \mu^{\epsilon'}.
\end{align}
The symmetric point $S_\mu$ is chosen in such a way that $p_i^2=\frac{3}{4}\mu^2$ and $p_i.p_j=-\mu^2/4$.
Using the relevant graphs of the Fig.~\ref{feyn_graphs} one can write
\begin{align}
\label{Gamma functions}
\Gamma^{(2)}(p,-p)  = Z|p|^\sigma &  \\
 -\frac{\lambda^2}{3!}\mu^{2\epsilon'}&\!\!\int \!\! \frac{\mathrm{d}^dq_1}{(2\pi)^d}\frac{\mathrm{d}^dq_2}{(2\pi)^d}\frac{1}{ |q_1|^\sigma|q_2|^\sigma|p-q_1-q_2|^\sigma}; \nn \\
\Gamma^{(4)}(p_1,p_2,p_3,p_4)|_{_{S_{\mu}}}&=\lambda \mu^{\epsilon'}Z_\lambda Z^2 \qquad \qquad \qquad\\
 -\frac{\lambda^2}{2}\frac{4!}{(2!)^3}\mu^{2\epsilon'}\!\! &\int\!\!\frac{\mathrm{d}^dq}{(2\pi)^d}\frac{1}{|q|^\sigma|p_1+p_2-q|^{\sigma}}. \nn
\end{align}
The integrals are infrared divergent for $\epsilon'>0$ and need to be calculated by analytical continuation from the
convergent region. This is the same situation as in the usual $\epsilon$-expansion in the short-range Ising model. Although the
integrals are complicated,  they can be calculated using the formulas in \cite{Itzykson},  and after using the renormalization conditions we will have
\begin{align} \label{Z and Zg}
Z&=1-\frac{\lambda^2}{3\sigma(4\pi)^d}\frac{\Gamma^3[\frac{d-\sigma}{2}]}
{\Gamma[\frac{3d-3\sigma}{2}]}\frac{\Gamma[\frac{3\sigma}{2}-d+1]}{\Gamma^3[\frac{\sigma}{2}]},\\
Z_\lambda&=1+\frac{3\lambda}{2(4\pi)^{d/2}} \frac{\Gamma^2[\frac{d-\sigma}{2}]}{\Gamma[d-\sigma]}\frac{\Gamma[\epsilon'/2]}{\Gamma^2[\sigma/2]}.
\end{align}
The very important point is that if we expand  $Z$ with respect to $\epsilon'$ there will not be any pole and one cannot
get sensible contribution to the critical exponent of the field $s(x)$. However, since the integrals are infrared
divergent, the right way to get a sensible perturbation theory is to expand $Z$ first around $\sigma=2$ and then around
$\epsilon'=0$. The situation is very similar to the short-range case;
 we have an integral which is divergent and would like to control its divergency. If we expand the above equations first around $\epsilon'$ we actually
get a finite term which is apparently wrong. Our choice of order of expansion is not arbitrary and 
it was actually forced by the divergent integrals. Since we have two parameters dimension $d$ and $\sigma$; and they can be changed independently, one can
first consider having $\delta\sigma=2-\sigma$ small and then do the perturbation theory with respect to $\epsilon'$.
After expanding $Z$ and $Z_g$ with respect to $\delta \sigma$ and then $\epsilon'$ we get
\begin{align} \label{Z and Zg2}
&Z=1\!-\!\frac{\lambda^2}{12(4\pi)^d\epsilon'}\!\!\left[\!1\!-\!\left(\!\frac{1}{2\epsilon'}\!+\!\frac{12\gamma\!-\!13}{8}\!\right)\!\delta\sigma\!+\dotsb\!\right]\!\!+\!\mathcal{O}\!\left(\lambda^3\right)\!,\\
&Z_\lambda=1+\frac{3\lambda}{(4\pi)^{d/2}\epsilon'}\left[1+(1-\gamma)\delta\sigma+\dotsb\right]\!+\mathcal{O}\!\left(\lambda^2\right)\!.
\end{align}
with $\gamma=0.5772...$ as the Euler-Mascheroni constant.
Using the Callan-Symanzik equation which states that the derivative of the bare quantities with respect to $\mu$ is zero, one can  get the beta functions as
\begin{align} \label{beta functions}
\beta(\lambda)=&\mu\frac{\partial }{\partial \mu}\lambda=-\epsilon'\lambda+\frac{3\lambda^2}{(4\pi)^{d/2}}\left[1\!+\!\left(1-\gamma\right)\delta\sigma\right]\!,\\
\gamma(\lambda)=&\mu\frac{\partial }{\partial \mu}\ln Z\!=\!\frac{\lambda^2}{6(4\pi)^d}\!\!\left[\!1\!-\!\!\left(\!\frac{1}{2\epsilon'}\!+\!\frac{12\gamma\!-\!13}{8}\!\right)\!\delta\sigma\!\right]\!\!.
\end{align}
To derive the above formula we first use the equation~(15) to get a relation between $\lambda$, $\lambda_0$ and $\mu$. Using the above beta functions at the critical point where $\beta(\lambda^{*})=0$ one can easily get the correction to the
mean-field value of the critical exponent $\delta\eta =\eta-(2-\sigma)$ as
\begin{equation} \label{deltaeta}
\delta\eta\!=\!\gamma(\lambda^{*})\!=\!\frac{1}{54}\epsilon'^2-\frac{1}{108}\epsilon'\delta\sigma-\frac{1}{432}\epsilon'^2(3-4\gamma)\delta\sigma+\dotsb
\end{equation}

\begin{figure} [htb]
\centering
\epsfxsize=240pt{\epsffile{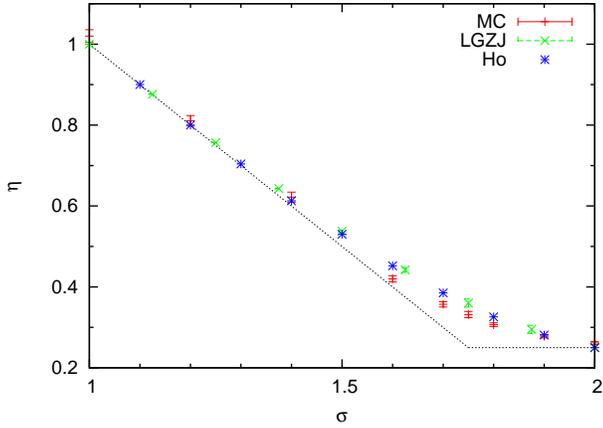}}
\caption{Data depicted in red are the result of the simulation in \cite{mp}. Green and blue data are coming from our
approximation using the results of \cite{LeGuillou} and \cite{Holovatch} respectively. 
The dotted line is Sak's result \cite{sak}.} \label{Figure:2}
\label{complete_graph}
\end{figure}

Based on our prescription it is obvious that in the $\epsilon'$-expansion of the $\eta$ exponent the zeroth order terms of $\delta\sigma$ expansion will be the same as the $\epsilon$-expansion of the short-range Ising model but with $\epsilon'=2\sigma-d$ instead of  $\epsilon=4-D$. So in principle close to the $\sigma=2$ we will have
\begin{equation}\label{eta exponents}
\eta=2-\sigma+\frac{1}{54}\epsilon'^2+\dotsb+\mathcal{O}(\delta\sigma),
\end{equation}
where dots represent the higher order terms of the $\epsilon'$-expansion. Since in our proposal of doing the RG we first
expand all the contributions around $\sigma=2$ and then around $\epsilon'=0$ we expect that the dots in the formula
(\ref{eta exponents}) are exactly the same  as in the short-range Ising model with $\epsilon'$ instead of $\epsilon$. The above expansion suggest that for $\delta\sigma$ small one can argue that the critical
exponent of the long-range Ising model in $d$ dimensions is approximately the same as  the critical exponent of
$D=4+d-2\sigma$ short-range Ising model. For the short-range Ising model $\delta\eta$ is known up to
$\epsilon^5$ for various dimensions \cite{LeGuillou,Holovatch}.  The first correction to this value comes from the second and
third terms of the equation (\ref{deltaeta}) which are both negative. If the higher order terms,
$\left(\delta\sigma\right)^n$ with $n\ge2$, do not change the sign of the contribution to $\eta$ one can conclude that
the critical exponent $\eta$ of the short-range Ising model in $D=4+d-2\sigma$ gives an upper bound for the $\delta\eta$
of the long-range Ising model in $d$ dimension. Of course this conjecture needs to be checked by calculating higher loop
corrections to the critical exponents. Based on the above arguments we compared in Fig.~\ref{complete_graph} the
$\eta$ coming from the numerical calculations for the long-range Ising model in two dimensions with the results
coming from the five loop calculation of $D=6-2\sigma$ dimensional short-range Ising model. The results are well
comparable in the region $1.75<\sigma<2$ and as we argued for the smaller values of $\sigma$ the actual values 
lie below our approximation.

\section{Conclusions}
In this letter we provided further numerical evidences that the long-range interaction
have an influence on the critical behavior of the Ising model for $\sigma \leq 2$, in contrast with
previous RG studies \cite{fisher,sak}.
We proposed a way to compute the influence on $\eta$ of a deviation from $\sigma=2$ based on renormalization group ideas. 
The main idea is the double expansion with respect to $\delta\sigma=2-\sigma$ and then $\epsilon'=2\sigma-d$ in a way that 
we get a non-trivial contribution to the wave function renormalization. Our analysis shows that close to  
$\sigma=2$ one can approximate the $\eta$ exponent of the  $d$ dimensional long-range Ising model with the same exponent of 
($d=4+D-2\sigma$)-dimensional short-range Ising model. 
Our results are in excellent agreement with the numerical results \cite{mp}.

\acknowledgments
We thank A. Gambassi, G. Gori, A. Trombettoni and  A. Codello for useful discussions. The work of M.~A.~Rajabpour was supported in part by FAPESP.

\end{document}